\documentclass[12pt]{article}
\usepackage{amsmath,amssymb,times,graphicx,mathdots}
\usepackage{color}
     \definecolor{darkred}{rgb}{0.75,0,0}
     \definecolor{darkgreen}{rgb}{0,0.5,0}
     \definecolor{darkblue}{rgb}{0,0,0.75}
     \definecolor{darkorange}{rgb}{1,0.9,0.1}		
\begin{document}
\pagestyle{plain}
\hsize = 6. in 				
\vsize = 8.5 in		
\hoffset = -0.3 in
\voffset = -0.5 in
\baselineskip = 0.26 in	

\def\vF{{\bf F}}
\def\vJ{{\bf J}}
\def\vX{{\bf X}}
\def\vecr{\mbox{\boldmath$x$}}
\def\vecv{\mbox{\boldmath$v$}}
\def\tf{{\widetilde{f}}}
\def\vv{{\bf{v}}}
\def\tv{{\widetilde{v}}}
\def\tu{{\widetilde{u}}}
\def\tp{{\widetilde{\rho}}}
\def\tR{{\widetilde{R}}}
\def\vx{\mbox{\boldmath$x$}}
\def\tx{\widetilde{x}}
\def\hat{\widehat{t}}
\def\ty{\widetilde{y}}
\def\mR{\mathbb{R}}
\def\mP{\mathbb{P}}
\def\mcA{\mathcal{A}}
\def\wtQ{\widetilde{\mathcal{Q}}}

\newcommand{\dab}[1]{\textcolor{darkblue}{#1}}
\newcommand{\jia}[1]{\textcolor{darkgreen}{#1}}
\newcommand{\bin}[1]{\textcolor{darkred}{#1}}
\newcommand{\daor}[1]{\textcolor{darkorange}{#1}}

\title{The Zeroth Law of Thermodynamics and 
Volume-Preserving Conservative System in
Equilibrium with Stochastic Damping}

\author{Hong Qian\footnote{
Email: qian@u.washington.edu,
Phone: 206-543-2584,
Fax: 206-685-1440.}\\[10pt]
Department of Applied Mathematics\\
University of Washington, Seattle\\
WA 98195-3925, U.S.A.}

\maketitle

\centerline{\bf Abstract}

We propose a mathematical formulation of the zeroth law of thermodynamics and develop a stochastic dynamical theory, with a consistent  irreversible thermodynamics, for systems possessing sustained conservative stationary current in phase space while in equilibrium with a heat bath.  The theory generalizes underdamped mechanical equilibrium: $dx=gdt+\{-D\nabla\phi dt+\sqrt{2D}dB(t)\}$, with $\nabla\cdot g=0$ and $\{\cdots\}$ respectively representing phase-volume preserving dynamics and stochastic damping. The zeroth law implies stationary distribution $u^{ss}(x)=e^{-\phi(x)}$.   We find an orthogonality $\nabla\phi\cdot g=0$ as a hallmark of the system.  Stochastic thermodynamics based on time reversal $\big(t,\phi,g\big)\rightarrow\big(-t,\phi,-g\big)$ is formulated: entropy production $e_p^{\#}(t)=-dF(t)/dt$; generalized ``heat'' $h_d^{\#}(t)=-dU(t)/dt$, $U(t)=\int_{\mathbb{R}^n} \phi(x)u(x,t)dx$ being ``internal energy'', and ``free energy'' $F(t)=U(t)+\int_{\mathbb{R}^n} u(x,t)\ln u(x,t)dx$  never increases.  Entropy follows $\frac{dS}{dt}=e_p^{\#}-h_d^{\#}$.  Our formulation is shown to be consistent with an earlier theory of P. Ao.  Its contradistinctions to other theories, potential-flux decomposition, stochastic Hamiltonian system with even and odd variables, Klein-Kramers equation, Freidlin-Wentzell's theory, and GENERIC,  are discussed.

\section{Introduction}
\label{sec1}

Newtonian deterministic dynamics of  particles and Gibbs' statistical 
treatment of heterogeneous equilibrium matters are two of the most 
important mathematical theories of physical phenomena in one's daily
experiences.  In recent years, there has been a significant progress in 
a mathematical theory of dynamics and thermodynamics of mesoscopic 
systems with a Markov-process description.  This theory is slowly
becoming a part of the dynamic branch of Gibbs' program which had 
started more than a century before.  

	The {\em stochastic thermodynamics} encompasses two 
types of nonequilibrium phenomena: one is a transient, non-stationary
stochastic processes approaching to an equilibrium, and another is
a stationary, ergodic nonequilibrium steady state (NESS) with 
sustained energy input balanced by dissipation
\cite{seifert_2,esposito_12,zqq,santillan_qian,ge_qian_pre_13,qian_decomp}.  
Taking stochastic fluctuations into account, the theory characterizes three 
distinct temporal behaviors: non-stationary transient relaxation, stationary fluctuating equilibrium with zero entropy production, and stationary fluctuating
NESS with positive entropy production.   The theory has already 
found natural applications in biochemical signal processing inside living cells \cite{gqq,tuyuhai,mehta-schwab}, free energy transduction in motors proteins \cite{chowdhury,ge_qian_pre_13}, and other molecular 
machines \cite{seifert_2}.  

	In mathematical terms, however, the definition of a ``stochastic 
equilibrium'' is far from universally agreed upon \cite{jqq}.  In Newtonian 
conservative dynamic theory, any system as a whole is automatically in 
an equilibrium, in a thermodynamic sense | there is simply no transient 
behavior, nor dissipation.  A transient phenomenon in this perspective only
occurs in a subsystem.  With the presence of ``frictions''
phenomenologically defined, the equilibrium
of a subsystem is the long-time behavior after 
damping with energy loss.   
However, a closer look indicates that there are two very different types 
of dynamic behavior in a mechanical equilibrium:  an overdamped
system with no oscillations and an underdamped system 
with continuous cyclic motion in $(p,q)$ phase space. 

	On the other hand, one of the major successes of stochastic 
dynamics is in the chemistry and biochemistry of single macromolecule 
in aqueous solution with overdamped dynamics, e.g., the Flory-Rouse-Zimm polymer theory \cite{doi_edward,schellman} and its NESS generalization \cite{qian_pre_02,qian_jpc_02}.   They are sometime 
collectively called soft matters
in physics \cite{smp}.  While overdamped stochastic dynamics is extensively studied with wide applications, underdamped systems with both {\em conservative oscillations} and {\em stochastic damping} \cite{wax_book, bergmann_lebowitz_1955, vankampen_57,ulhorn,graham_haken_71} have been studied only within physics community.  This includes, for examples, laser physics, electrical circuits, Josephson junctions, nanomechanical resonators, etc. \cite{mcclintock,dykman_chan}.  Many such systems exhibit stochastic resonance phenomena due to an interplay between stochastic dynamics and nonlinear oscillations \cite{hanggi_review,dykman_93,zqq}.

	All those above mentioned systems have a second-order dynamics
in which two types of variables, $p_i(t)$ and $q_i(t)$ are identified {\em a priori}. 
In a more abstract mathematical theory where variables 
$\{x_j(t)|j=1,2,\cdots\}$ are not distinguished, detailed balance has been 
a widely used mathematical criterion for equilibrium, for example in 
chemistry \cite{lewis_25,fowler_25,tolman_25}.  However, 
detailed balance condition in stochastic processes in fact 
depends upon identifying even and odd dynamic variables,
 a notion first proposed by Casimir in analogous to positions and velocities in classical mechanics \cite{casimir}.  In the context of Markov dynamics, they first appeared as $\alpha$ and $\beta$  variables  in the work of Machlup and Onsager \cite{machlup_onsager}, then studied by van Kampen \cite{vankampen_57}, Graham and Haken \cite{graham_haken_71}, and more recently in \cite{kim_qian_prl,kim_qian_pre,kwon_park_11,spinney_ford,munakata,gehao}.  For a brief summary of this matter, readers are referred to \cite{gardiner_book}.

	Furthermore, a canonical form of stochastic Hamiltonian dynamics with weak (under-)damping is a diffusion process represented by the Klein-Kramers equation, the Kolmogorov forward equation for a second-order Newton's equation of motion in a stochastic medium with frictions and random collisions.  As a system of stochastic differential equations (SDE), the diffusion matrix is singular in the Klein-Kramers equation.  Generalizing this approach encounters a mathematical difficulty due to the degeneracy of the diffusion:  The existence 
of an equilibrium ensemble is often difficult to establish with some rigor.\footnote{I thank Professor Min Qian of Peking University for extensive  discussions on stochastic processes, dynamical systems, and  time reversibility.  His 1979 paper \cite{mqian} in Chinese, which had inspired \cite{kim_qian_prl}, 
attemped to introduce a stochastic Hamiltonian system along the 
line of Klein, Kramers, Wang and Uhlenbeck, and illustrated the importance of fluctuation-dissipation relation  as a necessary condition for the
existence of Gibbs' canonical ensemble.}  

This paper presents a mathematical theory that generalizes underdamped equilibrium stochastic dynamics, together with an irreversible thermodynamics, whose applicability goes beyond traditional Hamiltonian systems \cite{andrey}.
Insteady of starting with a second-order stochastic differential equation 
and using detailed balance as the equilibrium constraint, we approach 
an underdamped stochastic dynamics from a rather
different starting point.  In nonlinear dynamical systems theory, a
conservative system with volume-preserving flow in phase space,
$\dot{x}=g(x)$ where $\nabla\cdot g(x)=0$,
is the natural generalization of Hamiltonian dynamics
\cite{andrey,strogatz,berger,gaspard}. 
%Ordinary differential equation 
%$\dot{x}=g(x)$ with $\nabla\cdot g(x)=0$ can also
%be represented as flow in phase space
%$\partial_tu(x,t)= -g(x)\cdot\nabla u(x,t)$. 
When coupled to an {\em stochastic damping with friction and
collisions}, it is natural to consider a stochastic differential 
equation (SDE)
\begin{equation}
         dx(t) = g(x)dt + \Big\{-D\nabla\phi(x)dt+ 
                 d\xi(t)\Big\}
\label{the_eq}
\end{equation}
in which the terms inside $\{\cdots\}$, in the spirit
of P. Langevin \cite{coffey_book}, represent the mean ``friction'' 
and rapid ``collision'' of the stochastic damping:
$\langle\xi(t)\xi(t')\rangle=2D\delta(t-t')$.    

	We focus on systems that approach to an equilibrium 
with stationary current. To introduce equilibrium condition into the 
stochastic setting, the present work departures from the traditional enforcing of fluctuation-dissipation relation to Eq. (\ref{the_eq}).  Instead, we generalize the idea of the  {\em zeroth law of thermodynamics} which states that two systems in equilibrum are essentially
unaltered whether they are in contact with or detached from 
each other:  We assume that the stationary 
density of the system (\ref{the_eq}), $u^{ss}(x)$, has to 
be the {\em same} with and without the damping part:
\begin{equation}
         u^{ss}(x) = e^{-\phi(x)}.
\label{new_fdr}
\end{equation}
Eq. (1) shows that its stationary state is invariant
under an additive constant $\phi(x)\rightarrow \phi(x)+C$.
The constant $C$ will be chosen such that the $u^{ss}(x)$
 in Eq. (2) is normalized.

Eq. (\ref{new_fdr}) generalized the notion of the zeroth law.
It shares the same spirit as detailed balance in statistical chemistry
\cite{lewis_25,fowler_25,tolman_25}, fluctuation-dissipation 
relation in statistical physics \cite{callen_welton_51,wyman_1,fdt_kubo}, 
and J. Wyman's thermodynamic linkage in macromolecular 
biochemistry \cite{wyman_2}: 
{\em If a bath is in equilibrium with a system, then
the equilibrium bath is invariant irrespective of whether  
it is in contact with the system or not.} 
A nonequilibirum steady state arises only when 
the condition in (\ref{new_fdr}) is violated due to
a dis-equilibrium between the bath and the conservative
dynamical system \cite{bergmann_lebowitz_1955,kim_qian_prl}.
Such a supposition underlies many of recent studies
on second-order (e.g., underdamped) stochastic dynamics
\cite{kim_qian_prl,kim_qian_pre,kwon_park_11,
spinney_ford,munakata,gehao};  its violation constitutes an
active driving force, e.g., a feedback control or a Maxwell's 
demon, from a nonequilibrium environment \cite{qian_arbp}. 

	 Introduced as above, an equilibrium stochastic system 
possesses an important orthogonal relation between 
$\nabla\phi$ and $g$.  To show this, we write the 
Kolmogorov forward equation for 
the SDE in (\ref{the_eq}):
\begin{equation}
   \frac{\partial  u(q,t)}{\partial t} = 
        \nabla\cdot\Big[D\nabla u - \Big(g(x)-D\nabla\phi(x)\Big)u
                   \Big].
\label{new_fpe}
\end{equation}
Then Eq.  \ref{new_fdr} implies 
$-\nabla\cdot\big[g(x)u^{ss}\big]
=u^{ss}(x)\big(g(x)\cdot\nabla\phi\big)=0$.
Therefore, for dynamics with $u^{ss}(x)>0$, the 
equilibrium condition implies the orthogonality. 
The stationary current is $g(x)u^{ss}(x)$.

	Sec. \ref{sec2} of the paper provides the mathematical basis
for a ``conservative stochastic thermodynamics'' in the presence 
of stationary probability current \cite{ao_ctp}.   Based on 
a new form of time reversal motivated by Hamiltonian 
dynamics, $\big(t,\phi,g\big)\rightarrow
\big(-t,\phi,-g\big)$ for Eq. \ref{the_eq} \cite{casimir}, 
a measure-theoretical entropy production $e_p^{\#}$ is 
introduced in Eq. \ref{def_ep}.  We show that it equals to the 
decreasing rate of a generalized free energy functional
$-dF/dt$, where \cite{ao_ctp} 
\begin{equation}
     F\big[u(x,t)\big] = \int_{\mathbb{R}^n} 
            u(x,t)\ln\Big(u(x,t)e^{\phi(x)}\Big) dx.
\end{equation}
$-dF/dt$, which is non-negative, has been called 
free energy dissipation in \cite{ge_qian_pre,feng_wang_11} 
and non-adiabatic entropy production in
\cite{sasa_prl,jarzynski,esposito}.  In the stationary state,
$e_p^{\#}\equiv 0$, and the probability current is analogous
to the inertia in mechanics and magnetic induction in electrical
circuits  \cite{mcclintock}.   

In 2004, P. Ao proposed a novel form of a non-detailed-balanced
stochastic process together with a decomposition scheme that yields 
stationary probability density and stationary current, as well 
as a steady state thermodynamics \cite{aoping_1,ao_ctp}:
\begin{subequations}
\begin{equation}
  \Big[S(x)+A(x)\Big]\frac{dx}{dt} = -\nabla\phi(x)
         +\zeta(x,t), \ \ \ x\in\mathbb{R}^n
\end{equation}
in which
\begin{equation}
             \langle \zeta(x,t)\zeta^T(x,t')\rangle = 2S(x)
           \delta(t-t').
\end{equation}
\label{ape}
\end{subequations}
One of the attractive features of Eq. (\ref{ape}a) is its
representation in terms of a symmetric matrix $S(x)$ and 
an antisymmetric matrix $A(x)$ along the stochastic path,
potentially facilitating computations for such stochastic systems
without detailed balance.   However the author has never
made the relationship clear between (\ref{ape}) and the 
general stochastic differential equation 
\begin{equation}
     dx(t)= b(x)dt+\Gamma(x)dw(t).
\label{the_sde_0}
\end{equation}
Ao's stochastic theory has 
intrigued and  mystified many researchers;
he and his coworkers have further carried out 
explicit computations for linear stochatsic differential 
equations following the general theory  \cite{kat_pnas}.
One of the key features in the linear system is an 
orthogonality between the gradient of stationary density 
and the stationary probability current. 
In Sec. \ref{sec3}, we shall show that  the system in 
Eq. \ref{ape} is consistent with the stochastic dynamics 
with volume-preserving flow introduced in 
Eqs. \ref{the_eq}, \ref{new_fdr}, and \ref{new_fpe}.   

The theory we present is a synthesis of several known results. 
To clarify, we shall reiterate its novelty:  It ($i$)
develops a underdamped stochastic dynamics in the general term 
of dynamical systems without the need of a Hamiltonian, nor the 
identification of even and odd variables.  The classical Klein-Kramers
equation for stochastic Newtonian dynamics is simply a special case
(see Sec. \ref{sec4.1}).  ($ii$) mathematically  
formulated the zeroth law of thermodynamics as 
an equilibrium condition; ($iii$) derived an orthogonal relation 
between gradient $\nabla\phi$ and current $g$.  Note this 
orthogonality is not the same as that obained in the Freidlin-Wentzell 
theory \cite{fw_70,graham_tel}, where the term corresponding to $g$ 
was not a divergence-free current in general (see Sec. \ref{sec.5.1}).  
($iv$) shows a consistency with the stochastic dynamical equation 
proposed by Ao in \cite{aoping_1};  ($v$) introduced a trajectory-based
entropy production formula using the time reversal
$(t,\phi,g)\rightarrow (-t,\phi,-g)$ and derived irreversible thermodynamic
equations (\ref{2ndlaw}) and (\ref{eq_4_dsdt}) for the thermodynamics.

\section{Stochastic thermodynamics}  
\label{sec2}

\begin{figure}[htb]
\begin{center}
\includegraphics[width=4.5in]{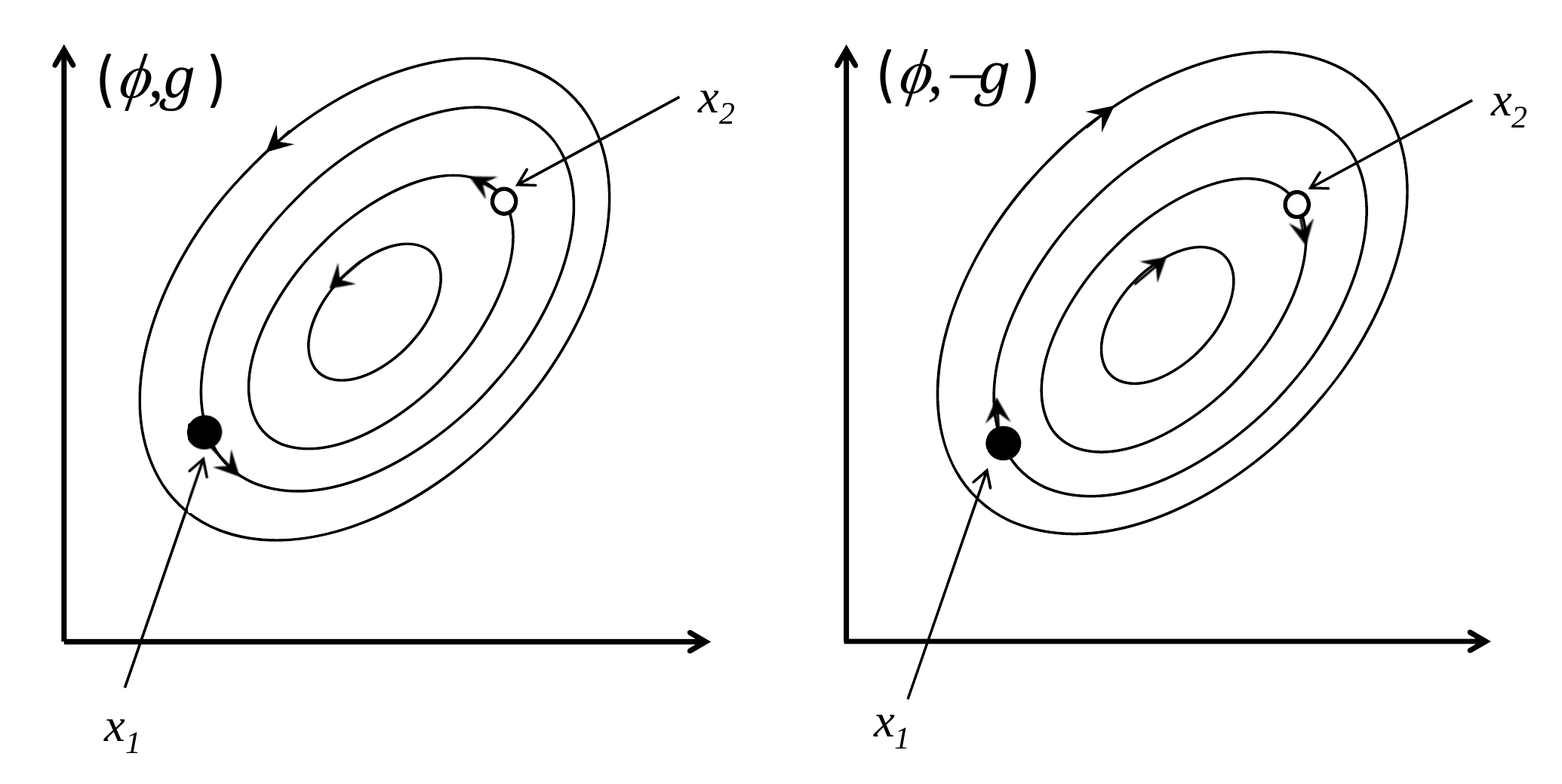}
\end{center}
\caption{
In stationarity with $\tau\ge t$, the joint probability distribution
$\Pr\big\{X(t)=x_1,X(\tau)=x_2\big\}$ for the diffusion process on the left,
according (\ref{the_eq}) with $\big(\phi,g\big)$, {\em i.e.},  
$dX(t)=\big(g-D\nabla\phi\big)dt+d\xi(t)$ and
$\nabla\phi\cdot g=0$, is the same as the joint probability distribution 
$\Pr\big\{X^-(t)=x_2;X^-(\tau)=x_1\big\}$ for the diffusion process on the 
right, $X^-(t)$ with $\big(\phi,-g\big)$,
under a time reversals: $\Pr\big\{X^-(-t)=x_2;X^-(-\tau)=x_1\big\}$ $=$
$\Pr\big\{X^-(\tau-t)=x_2;X^-(0)=x_1\big\}$ $=$ 
$\Pr\big\{X(0)=x_1,X(\tau-t)=x_2\big\}$ $=$ $\Pr\big\{X(t)=x_1;X(\tau)=x_2\big\}$.
The closed cycles in both plots, the contours of $\phi$, are the same; the 
actual speed along a contour is determined by $\|g\|$.  
}
\label{fig}
\end{figure}

In a Hamiltonian system $\dot{q} =\partial H/\partial p$
and  $\dot{p} =-\partial H/\partial q$, transformation $(q,p) \rightarrow (q,-p)$ 
is equivalent to transforming $H\rightarrow -H$ \cite{strogatz,berger,gaspard}. 
In the theory of diffusion process, the Kolmogorov forward equation for 
SDE (\ref{the_sde_0}) with stationary density $u^{ss}(x)$ is
\begin{equation}
       \frac{\partial u(x,t)}{\partial t} = \nabla\cdot\left(A(x)u^{ss}(x)
                  \nabla \left(\frac{u(x,t)}{u^{ss}(x)}\right) - 
              \Big\{b(x)-A(x)\nabla u^{ss}(x)\Big\} u(x,t)\right),
\end{equation}
in which $ A(x) = \frac{1}{2}\Gamma(x)\Gamma^T(x)$.  
The time-reversed process to (\ref{the_sde_0})
satisfies the Kolmogorov equation with a sign-change 
for the divergence-free term in $\{ \cdots \}$ \cite{qian_decomp}.  
These observations suggest that  a meaningful
time-reversed process for SDE in (\ref{the_eq}) is 
$(x,g)\rightarrow (x,-g)$, as illustrated in 
Fig. \ref{fig}. The time reversal tranformation then is 
$(t,x,g)\rightarrow (-t,x,-g)$: {\em The time-reversed process
for time-reversed stochastic path.}   Also see Sec. \ref{sec.4.1} 
for more justifications.  

In terms of such a time reversal, let us consider a 
stochatsic path $\omega_{[t,\tau]}=\big\{x(s)\big|t\le s\le\tau\big\}$
and its time-reversed path 
$r\omega_{[t,\tau]}=\big\{x(t+\tau-s)\big|t\le s\le \tau\big\}$
under the probability measures $\mathbb{P}^{+g}$
and $\mathbb{P}^{-g}$, rspectively, defined by SDEs 
$dX=\big(g-D\nabla\phi\big)dt+d\xi(t)$ and its 
time-reversal $dX^-=\big(-g-D\nabla\phi\big)dt+d\xi(t)$.
Denoting $\mathbb{P}^{-}\big(\omega_{[t,\tau]}\big)$ $\equiv$ 
$\mathbb{P}^{-g}\big(r\omega_{[t,\tau]}\big)$, 
a sample-path based entropy production can be 
introduced \cite{qian_decomp,esposito,jqq,qq_85,crooks,seifert_05,gj_07}:
\begin{eqnarray}
   e_p^{\#}(t) &=& \lim_{\tau\rightarrow t}E^{\mathbb{P}^{+g}}\left[
           \frac{1}{|\tau-t|}
              \ln\frac{d\mathbb{P}^{+g}}
          {d\mathbb{P}^-}\big(\omega_{[t,\tau]}\big)\right]
\label{def_ep}
\\[6pt]
       &=& -\int_{\mathbb{R}^n} J(x,t)\cdot
                  \nabla\ln
          \left(\frac{u(x,t)}{u^{eq}(x)}\right) dx,
\\[6pt]
      &=& \int_{\mathbb{R}^n}
       u(x,t)\nabla\ln\Big( u(x,t)e^{\phi(x)}\Big)\cdot D
                  \nabla\ln
          \Big(u(x,t)e^{\phi(x)}\Big) dx \ \ge\ 0,
\end{eqnarray}
where $J(x,t)=\big(g-D\nabla\phi\big)u(x,t)-D\nabla u(x,t)$,
and $E^{\mathbb{P}^{+g}}\big[\cdots\big]$ denotes
ensemble everage with respect to $\mathbb{P}^{+g}$.
We note that the $g$ term has disappeared in the
final expression: Entropy production is purely 
determined by the damping mechanism.  Note that we 
have changed the notation $u^{ss}(x)$ to $u^{eq}(x)$
to emphasize that the steady state is an equilibrium with
conservative current. 

It is important to point out that the $e_p^{\#}$ introdiced in
Eq. \ref{def_ep} is different from the standard
entropy production for a stochastic diffusion
\cite{jqq,seifert_05,gj_07,qian_decomp}.  In the 
literature, $e_p^{\#}$ has been called free energy 
dissipation or non-adiabatic entropy production
\cite{ge_qian_pre,sasa_prl,jarzynski,esposito}.

One can also define a generalized 
nonequilibrium free energy \cite{qian_decomp},
\begin{equation}
    F(t) = \int_{\mathbb{R}^n} u(x,t)\ln
          \left(\frac{u(x,t)}{u^{eq}(x)}\right)\ dx.
\label{defft}
\end{equation}
Then in terms of the $e_p^{\#}$, it has been shown 
in \cite{ao_ctp} that:
\begin{equation}
     \frac{dF(t)}{dt} =  \int_{\mathbb{R}^n} J(x,t)\cdot
            \nabla\ln\left(\frac{u(x,t)}{u^{eq}(x)}\right)
                    \ dx \ = \ - e_p^{\#}(t)  \le  0.
\label{2ndlaw}
\end{equation}
Eq. \ref{2ndlaw} should be interpreted as follows:
{\em For Gibbs' canonical ensemble with 
non-uniform $u^{eq}(x) = e^{-\phi(x)}$, the
thermodynamic potential is free energy $F$ 
whose decreasing rate is the entropy production $e_p^{\#}$.}

The generalized nonequilibrium free energy $F(t)$ 
in (\ref{defft}) can be decomposed into $U(t)-S(t)$ with
\begin{equation}
   U(t) = -\int_{\mathbb{R}^n} u(x,t)\ln
          u^{eq}(x)\ dx,  \  \
  S(t) = -\int_{\mathbb{R}^n} u(x,t)\ln
          u(x,t)\ dx.
\label{eq0013}
\end{equation}
They are interpreted as internal (conservative) energy
and entropy, respectively.  Furthermore, 
\begin{equation}
    \frac{dU}{dt} = -\frac{d}{dt}\left(\int_{\mathbb{R}^n}
     u(x,t)\ln u^{eq}(x) dx\right) 
   =  \int_{\mathbb{R}^n} 
     \nabla\phi\cdot D\nabla\ln\Big(u(x,t)
            e^{\phi}\Big)\ dx .
\label{heat}
\end{equation}
We note again that the $g$ term has completely 
disappeared in Eqs. (\ref{eq0013}) and (\ref{heat}). 
The right-hand-side of (\ref{heat}) can be interpreted 
as ``heat flux'', analogous to the
heat in classical thermodynamics, $h_d^{\#}(t)$.  Then,
\begin{equation}
  \frac{dS}{dt} -  e_p^{\#}(t) = - h_d^{\#}(t) = \frac{dU}{dt},
\label{eq_4_dsdt}
\end{equation}  
which is precisely the entropy balance equation in
Dutch school of nonequilibrium thermodynamics 
\cite{degroot}: $\frac{dS}{dt}=\frac{d_iS}{dt}
+\frac{d_eS}{dt}$.  See also \cite{bergmann_lebowitz_1955}
and \cite{ge_qian_pre_13} for discussions on the 
meanings of these terms as the system's entropy change, 
total entropy prodction, and heat dissipation. 

In mathematics, the functional $F\big[u(t)\big]$, as the 
Gibbs-Shannon entropy with respect to a 
non-uniform probability measure with density $u^{eq}$, 
is known to contain two {\em a priori} estimates, based 
on respectively the non-negativity of $F$ and 
$e_p^{\#}$ \cite{villani_08}.

\section{Conservative dynamics with 
stationary current and stochastic damping}
\label{sec3}

	The previous section has established a 
self-consistent underdamped stochastic thermodynamics
with conservative stationary current.
We now show that the stochastic dynamics defined in 
Eqs. (\ref{the_eq}), (\ref{new_fdr}), and (\ref{new_fpe}) 
with orthogonal $\nabla\phi$ and $g$, is consistent with 
Ao's model (\ref{ape}).

Without being 
mathematically rigorous, one can formally establish the 
relation between Eq. \ref{ape} and the conventional SDE (\ref{the_sde_0}):  
Introducing a transformation via an auxiliary matrix inversion 
$\big(S(x)+A(x)\big)^{-1}$
$=$ $G(x)$ one obtaines \cite{aoping_1}
\begin{equation}
   \frac{dx}{dt} = -G(x)\nabla\phi(x)+ \xi(x,t),
\label{the_sde}
\end{equation}
with
\begin{eqnarray}
   \big\langle \xi(x,t)\xi^T(x,t')\big\rangle &=& 
   \Big\langle G(x)\zeta(x,t)\zeta^T(x,t')
                G(x)^T\Big\rangle 
\nonumber\\[5pt]
   &=& 2G(x)\ S(x)\ G^{T}(x)\ \delta(t-t').
\end{eqnarray}  
Then the associated Kolmogorov partial differential equation 
in divergence form, following \cite{yin_ao_06,yuan_ao}, is,
\begin{equation}
  \frac{\partial u(x,t)}{dt} = \nabla\cdot
       \Big[G(x)S(x)G^T(x)
          \nabla u + G(x)\nabla\phi(x)u \Big].
\end{equation}

One of the key results in \cite{aoping_1} is that the
stationary density for the stochastic process being
$e^{-\phi(x)}$.  Then the stationary current $J(x)$  
satisfies \cite{wangpnas,feng_wang_11}
\begin{equation}
           J(x) e^{\phi(x)} = G(x)S(x)G^T(x)
          \nabla\phi(x)-G(x)\nabla\phi(x).
\label{the_key_eq}
\end{equation}
From Eq. \ref{the_key_eq} one has
\begin{eqnarray}
   \big(\nabla\phi(x)\big)^T\cdot J(x)  e^{\phi(x)}
                &=& \big(\nabla\phi(x)\big)^T \cdot
    \Big[GSG^T \nabla\phi(x)-G\nabla\phi(x)\Big]
\nonumber\\[5pt]
    &=&  \Big(G^T\nabla\phi(x)\Big)^T\cdot\Big[
                     -S+G^{-T}\Big] 
                    \Big(G^{T}\nabla\phi(x)\Big)
\nonumber\\[5pt]
    &=&  \Big(G^T\nabla\phi(x)\Big)^T\cdot\Big[
                     -S+S+A^T\Big] 
                    \Big(G^{T}\nabla\phi(x)\Big)
\nonumber\\[5pt]
    &=& 0.
\label{ortho}
\end{eqnarray}
Since $e^{\phi(x)}\neq 0$, Eq. \ref{ortho} means 
$\big(\nabla\phi(x)\big)^T\cdot J(x)=0$ is a 
 necessary condition for Ao's stochastic model.
It is a special case of the 
process introduced in Sec. \ref{sec1}.

\section{Relations to other theories}

	An oscillatory motion in an overdamped mechanical 
system (e.g., a macromolecule in
an aqueous solution) is due to external drive with dissipation; 
while an oscillatory motion in an underdamped system is 
a part of conservative dynamics.  How to develop a 
mathematical thermodynamic theory for 
stochastic systems with stationary current, therefore, is a 
fundamental issue of mesoscopic dynamics \cite{qian_decomp}.
Because of its centrality, there have been many theories
that are relevant to the present work.   We shall discuss several
of them that we have studied.

\subsection{Klein-Kramers equation and stochastic Hamiltonian 
dynamics}
\label{sec4.1}

Following the seminal work of Langevin, Klein and Kramers, it
is a custom to writes a stochastic Newtonian dynamics for a 
subsystem of 1-degree of freedom, with damping, as:
\begin{equation}
                \left\{ \begin{array}{l} 
               dx  =  ydt, \\[5pt]
               mdy =  -\left[
                    \frac{dU(x)}{dx} + \eta(x) y \right] dt 
                    + \sqrt{k_BT\eta(x)} dB(t),  
                  \end{array}\right.
\label{21}
\end{equation}	
in which the coefficient of $dB(t)$ is fixed by
Einstein's relation, which guarantees Maxwell-Boltzmann
stationary distribution $e^{-H(x,y)/k_BT}$ with $H(x,y)=\frac{1}{2}my^2
+U(x)$.  In the notations of our Eq. (\ref{the_eq}):
\begin{equation}
      \vec{g} = \left(\begin{array}{c}
                          y  \\[5pt]   -\frac{1}{m}U'(x) \end{array}\right)
                   = \frac{1}{m}\left(\begin{array}{cc}
                               0 & 1 \\[5pt]  -1 & 0 \end{array}\right)\nabla H,
\end{equation}
\begin{equation}
         \nabla\cdot \vec{g}=  \frac{\partial}{\partial x} y
                  -\frac{\partial}{\partial y} \left(\frac{U'(x)}{m}\right) = 0, 
\end{equation}
\begin{equation}
     \phi = \frac{1}{k_BT}\left(\frac{my^2}{2}+U(x)\right),  \  \
               \nabla\phi = \frac{1}{k_BT}\Big(U'(x), my\Big), 
\end{equation}
therefore, $\vec{g}\cdot\nabla\phi = 0$.

	In fact, one has a more general stochastic 
Hamiltonian system \cite{mqian,kim_qian_prl}:
\begin{equation}
   \frac{d}{dt}\left(\begin{array}{c}
           x \\[5pt] y \end{array}\right)
        = \left(\begin{array}{cc}
            \frac{\partial H}{\partial y}+\eta_x(x,y)
            \\[8pt] 
              -\frac{\partial H}{\partial x} + \eta_y(x,y)
           \end{array}\right)+ 
          \Gamma(x,y)\left(\begin{array}{c}
             \frac{dw_x}{dt} \\[8pt] 
             \frac{dw_y}{dt}
            \end{array}\right),
\label{shs}
\end{equation}
in which both $x$ and $y$ are $n$-dimensional
vectors, and $\Gamma$ is a $2n\times 2n$ matrix.
Then, following our formalism, the $2n$ vector
\begin{equation}
     \vec{g} = \left(\begin{array}{c}
                \frac{\partial H}{\partial y} \\[6pt] 
                 -\frac{\partial H}{\partial x}
                     \end{array}
                   \right),
\end{equation}
$\phi = -\ln u^{eq}$, and 
\begin{equation}
      \left[\frac{1}{2}\Big(
      \big(\nabla_x, \nabla_y \big)
      \Gamma(x,y)\Gamma^T(x,y)\Big)^T -
      \left(\begin{array}{c}
              \eta_x(x,y) \\ \eta_y(x,y) \end{array}
          \right)\right]u^{eq}(x,y) = 0.
\label{eq_0140}
\end{equation}
Using (\ref{eq_0140}), the Fokker-Planck equation 
for (\ref{shs}),
\begin{equation}
    \frac{\partial u}{\partial t}
    = \big(\nabla_x, \nabla_y\big)\left[\frac{1}{2}
      \Big(\big(\nabla_x, \nabla_y \big)
      \Gamma(x,y)\Gamma^T(x,y)\Big)^T-
      \left(\begin{array}{c}
             \frac{\partial H}{\partial y}+\eta_x \\[8pt] 
       -\frac{\partial H}{\partial x}+\eta_y \end{array}
          \right)\right] u,
\label{eq_010}
\end{equation}
can be written as 
\begin{equation}
   \frac{\partial u}{\partial t}
    = \big(\nabla_x, \nabla_y\big)\left\{\frac{1}{2}\Gamma(x,y)\Gamma^T(x,y)
      \left[
      \left(\begin{array}{c}
                 \nabla_x \\[6pt] \nabla_y \end{array}
           \right)-\left(\begin{array}{c}
                 \nabla_x\ln u^{eq} \\[6pt] \nabla_y\ln u^{eq} 
           \end{array}\right)\right]-
      \left(\begin{array}{c}
             \frac{\partial H}{\partial y}\\[8pt] 
       -\frac{\partial H}{\partial x} \end{array}
          \right)\right\} u.
\label{0270}
\end{equation}
It is easy to verify that the orthogonal relation 
$\vec{g} \cdot \nabla\phi = 0$ guarantees
$e^{-H(x,y)}$ being the equilibrium solution to (\ref{0270}).
This yields a fluctuation-dissipation relation like equation
for stochastic damping:
\begin{equation}
              \left(\begin{array}{c}
              \eta_x \\ \eta_y \end{array}
          \right) = \frac{1}{2}e^{H(x,y)} \Big(
      \big(\nabla_x, \nabla_y \big)
      \Gamma(x,y)\Gamma^T(x,y)\Big)^Te^{-H(x,y)}.
\end{equation}
Newtonian dynamics is a special case with singular 
$\Gamma$ \cite{kim_qian_prl}
\begin{equation}
         \Gamma(x,y)\Gamma^T(x,y) = \left(
               \begin{array}{cc}
                     0 & 0 \\   0 &  1  \end{array}\right), \
              \left(\begin{array}{c}
              \eta_x \\ \eta_y \end{array} \right) =    \left(\begin{array}{c}
              0 \\  -\nabla_y H(x,y) \end{array} \right). 
\end{equation}

{\bf\em It\={o} vs. divergence form.}
There are several mathematical choices for the 
intergration of an SDE with multiplicative 
noise,  {\em e.g.}, It\={o}, Stratonovich, or Ao's 
divergence form \cite{yuan_ao}.  We note that
we started with It\={o}'s convention in Eq. (\ref{eq_010}).  
However withthe the zeroth law, the resulting  Eq. (\ref{0270}) 
in fact is in the divergence form.  The final
partial differential equation in fact is independent of
the choice of stochastic intergration.  To see this more 
clearly, consider the Fokker-Planck equation for 
$dx=\big(g(x)+\eta(x)\big)dt+\sqrt{2D(x)}dB(t)$,
in It\={o}'s form: 
\begin{equation}
   \frac{\partial u(x,t)}{\partial t}
   = \nabla\cdot\Big[\Big(\nabla\big(D(x)u(x,t)\big)
                       -\eta(x)u(x,t)\Big) 
                   -g(x)u(x,t)\Big],
\label{eq_a}
\end{equation}
with $\eta(x)$ and $\phi(x)$ being related via
\begin{equation}
      \nabla\Big(D(x)e^{-\phi(x)}\Big)-\eta(x)e^{-\phi(x)} = 0.
\label{eq_b}
\end{equation}
Solving $\eta(x)$ from Eq. (\ref{eq_b}) and substituting
it into (\ref{eq_a}), we have
\begin{equation}
   \frac{\partial u(x,t)}{\partial t}
   = \nabla\cdot\Big[D(x)\Big(\nabla u(x,t)
               +u(x,t)\nabla\phi(x)\Big) 
                   -g(x)u(x,t)\Big].
\label{new_kke}
\end{equation}

\subsection{Wang's Hodge-like decomposition}

The orthogonality in Eq. \ref{ortho} leads to several
interesting properties for the stochastic dynamics.
First, noting the SDE in (\ref{the_sde_0})  and the 
relation in (\ref{the_key_eq}), we have the drift
\begin{equation}
    b(x) = -G(x)\nabla\phi(x) \ =\ 
        -D\nabla\phi(x) +  J(x) e^{\phi(x)},
\label{wang_decomp}
\end{equation}
and $\langle\xi(t)\xi(t')\rangle = 2D\delta\big(t-t'\big)$.
Denoting $g(x)=J(x) e^{\phi(x)}$, then
\begin{equation}
            b(x)=-D\nabla\phi+g,\  
              \nabla\cdot g = 0, \ 
              \nabla\phi\cdot g = 0.
\label{hodgeplus}
\end{equation}  
The right-hand-side of (\ref{wang_decomp}) are Wang's 
potential and flux landscapes \cite{wangpnas,feng_wang_11} 
for a general SDE.  The orthogonality between the gradient 
and current terms is an additional feature of Ao's
stochastic processes.   The first two equations in
(\ref{hodgeplus}) are a Helmholtz-Hodge-like 
decomposition with diffusion matrix $D$ \cite{qw_cmp}.  
As far as we know, there is no orthogonality in a Hodge
decomposition in general.

{\bf\em Xing's Hamiltonian representation.}
For a Hamiltonian system,  the orthogonality is a
consequence of a damped Hamiltonian dynamics
\cite{bergmann_lebowitz_1955,mqian} in equilibrium with 
detailed-balanced stochastic fluctuations. 
Then the stationary process defined by (\ref{eq_010}) 
has a distribution $u^{eq}=e^{-H}$ as well as 
a conservative rotation in phase space: 
\begin{equation}
       J(x,y) = -\left(\begin{array}{c}
             \frac{\partial H}{\partial y} \\[8pt] 
       -\frac{\partial H}{\partial x}\end{array}
          \right) u^{eq}(x,y). 
\label{eq_0120}
\end{equation}
Indeed, $\nabla u^{eq}\cdot J = 0$.  
Eqs. (\ref{eq_0140}) and (\ref{eq_010}) are 
based on It\={o}'s integration.  It is easy to see that
if another convention for stochastic integration is chosen, both
equations will have different expressions; but Eqs. (\ref{0270})
and (\ref{eq_0120}) are invariant.

	For any SDE (\ref{the_sde_0}) with stationary 
density $e^{-\phi}$ and flux $J$, if $\nabla\phi\cdot J=0$, 
then the SDE can be re-written
as
\begin{equation}
   dx(t) =   \Big\{-D\nabla\phi(x)+d\xi(t)\Big\}+g(x)
\end{equation} 
in which the first two terms in the $\{\cdots\}$ can be ``interpreted'' as
the heat bath with detailed balance.  They are analogous to the 
mean ``friction'' and rapid ``collision'' in classical mechanics, defining the
notions of dissipation and fluctuation in a general stochastic 
dynamics.   The dynamics described by the 
$g(x)$, on the other hand, is a deterministic conservative system \cite{strogatz}
with a volume preserving flow in phase space \cite{andrey}: 
\begin{equation}
   \frac{d}{dt}\int_{\mathfrak{D}(t)} dx = 
         \oint_{\partial\mathfrak{D}} g(x)\cdot d\vec{S}
          = \int_{\mathfrak{D}}
              \nabla\cdot g(x)\ dx = 0.
\end{equation}
Therefore, it is not surprising that Xing is able to represent
Ao's process mathematically as a very large, conservative
Hamiltonian system in which stochastic damping corresponds to 
a harmonic bath \cite{xing}.  One should not confuse this result,
however, with the Hamiltonian dynamics that defines 
conditional most probable path in a diffusion process \cite{wang_jcp_10,ge_qian_ldt}.
The relation between these two Hamiltonian systems, if any,  remains 
to be elucidated.

{\bf\em Grmela-\"{O}ttinger's GENERIC.}  
One can also re-write the general stochastic Hamiltonian 
system Eq. (\ref{shs})  as
\begin{equation}
   \left(\begin{array}{c}
    dx \\[6pt] dy \end{array}\right)
     = \left[\left(\begin{array}{c}
            \frac{\partial H}{\partial y} \\[8pt]
           -\frac{\partial H}{\partial x} \end{array}
       \right) - \left(\begin{array}{cc}
              D_{xx} & D_{xy} \\[7pt] D_{yx} & D_{yy}
             \end{array}\right) 
          \left(\begin{array}{c}
            \frac{\partial\phi}{\partial x} \\[8pt]
           \frac{\partial\phi}{\partial y} \end{array}\right)
           \right] dt +  \left(\begin{array}{c}
    d\xi_x(t) \\[6pt] d\xi_y(t) \end{array}\right).
\end{equation}
The deterministic part here has the GENERIC form proposed 
by Grmela and \"{O}ttinger \cite{ottinger}.
The orthogonality has also figured prominently in 
the GENERIC structure which has a rich geometric
interpretation.

\subsection{Detailed balance in systems with even and odd variables}
\label{sec.4.1}

There have been extensive discussions on detailed balance in stochastic 
differential equation with even and odd variables.  See earlier work 
\cite{machlup_onsager,bergmann_lebowitz_1955,vankampen_57,ulhorn,graham_haken_71},
a nice summary in the textbook \cite{gardiner_book}, and more
recent papers \cite{kim_qian_prl,kim_qian_pre,kwon_park_11,kwon_ao_11,spinney_ford,munakata,gehao}.  Detailed balance for system with position 
$\vecr$ and velocity $\vecv$ is defined through a symmetry
between the transtion
$\big(\vecr,\vecv,t\big)\rightarrow \big(\vecr',\vecv',t+\tau\big)$ and 
transition
$\big(\vecr',-\vecv',t\big)\rightarrow \big(\vecr,-\vecv',t+\tau\big)$.  
The {\em detailed balance condition} in an even-and-odd 
system is shown to be sufficient for the probability current 
$\vJ=-\vJ^-$ where the $\vJ^-$ is the stationary current under time 
reversal (\cite{gardiner_book}, Eq. 5.3.53).  Also, for constant 
diffusion matrix, the cross terms between even and odd variables
are necessarily zero. Furthermore, the stationary solution is ``solved''
in \cite{gardiner_book} (Eq. 5.3.85).  Finally, recognizing the orthogonal
condition presented in the present work, their Eq. 5.3.82 can
be simplied into $\sum_i\frac{\partial}{\partial x_i}I_i(\vx)=0$;
where $I$ is our $g$.

\subsection{Temperature  and NESS systems in spatial contact}

	There have been continous efforts to introduce the concept 
of temperature into systems in NESS \cite{zero1,zero2,zero3}. 
Most of these work focused on stochastic interacting particle systems, 
in which a temperature difference is natually defined as a spatial gradient.
Two approaches have been employed: ($a$) Legendre transform
in conjunction with an energy conservation, and ($b$) 
empirically establishing an intensive quantity via numerical 
computations.   All these approaches are essentially ``thermodynamic''
in nature, while the zeroth law in the present work is formulated as a
statement about stationary distributions.  This is a much stronger  
consition.  It is also applicable to chemical potential equilibration 
as well as temperature equilibration.

We note, however, that our orthogonality
$\nabla\phi \cdot g=0$ is effectively a conservation law;
therefore it will be interesting to explore the possibility of
 introducing an conjugate intensive quantity via Legendre
transform.  Also, the fact that any linear stochastic dynamical
system automatically satisfies $\nabla\phi\cdot g=0$
\cite{kat_pnas} seems to sugget that a pseudo-temperature 
could be defined for system with linear irreversibility. 
\vskip 0.5cm

\subsection{Three different types of time reversal}

	We now consider the general SDE in Eq. \ref{the_sde_0}.  
In \cite{qian_decomp}, we have introduced the
notion of a {\em canonical conservative dynamics} with
respect to a differentiable invariant density $u^{ss}(x)$:
$\dot{x}=j(x)$ with $\nabla\cdot\big(u^{ss}(x)j(x)\big)=0$.
The general SDE,  when its stationary
density is known, can always be re-written 
as \cite{qian_decomp,gardiner_book}
\begin{equation}
  dx(t) = \Big(b(x)+D(x)\nabla\phi(x)\Big)dt 
         + \Big\{-D(x)\nabla\phi(x)\ dt+d\xi(t)\Big\},
\label{can_con}
\end{equation}
in which $e^{-\phi(x)}=u^{ss}(x)$ is the stationary
density.  Then $j(x)=b(x)+D(x)\nabla\ln\phi(x)$ is 
a canonical conservative dynamics with respect to $u^{ss}(x)$.
\cite{qian_decomp} also shows that under
time reveral $\big(t,\phi,j\big)\rightarrow \big(-t,\phi,-j\big)$,
the system in (\ref{can_con}) has again entropy production 
$e_p^{\#}(t)=-dF/dt$.
 
	One can further decompose the term $j(x)$ into 
parallel and perpendicular to $\nabla\phi(x)$:
$j(x) = j_{\parallel}(x) + j_{\perp}(x)$, with
\begin{equation}
       j_{\parallel}(x) = \left(\frac{j(x)\cdot\nabla\phi(x)}
                  {\|\nabla\phi(x)\|^2}\right)\nabla\phi(x),  \ \
       j_{\perp}(x) = j(x) -  j_{\parallel}(x),
\end{equation}
where $\nabla\phi(x)\cdot j_{\perp}(x)=0$.  Then
\begin{equation}
    b(x) = -D(x)\nabla\phi(x)+j_{\perp}(x)
                 +j_{\parallel}(x).
\label{3j}
\end{equation}
The last term can also be written as
\begin{equation}
     j_{\parallel}(x) = \left(\frac{\nabla\cdot j(x)}
                  {\|\nabla\phi(x)\|^2}\right)\nabla\phi(x).
\end{equation}
It represents the non-conservative nature of 
$j(x)$ \cite{andrey}.  According to time reversal
$\big(t,\phi,j\big)\rightarrow\big(-t,\phi,j\big)$, all $j(x)$
contributes to stationary entropy production \cite{jqq,zqq}; 
and according to time reversal 
$\big(t,\phi,j\big)\rightarrow\big(-t,\phi,-j\big)$, there will be no 
stationary entropy production \cite{qian_decomp}.
The present work and Eq. \ref{3j} suggests yet another time
reversal:
$\big(t,\phi,j_{\perp}\big)\rightarrow\big(-t,\phi,-j_{\perp}\big)$,
under which stationary dissipation arises from $j_{\parallel}$.  
These three different time reversals reflect assumptions
based on over-damped, non-damped or under-damped nature of
a stationary dynamics of a subsystem.   For systems with 
overdamped time reversibility, they have a potential
condition $D^{-1}(x)b(x)=\nabla\phi$ and the stationary density
is $e^{-\phi(x)}$ with zero stationary current. For systems with
underdamped time reversibility, they have an orthogonal
decompositon $b(x)=D(x)\nabla\phi+g$
with $\nabla\times g=0$ and $g\cdot D(x)\nabla\phi=0$.
Then the stationary density and current are $e^{-\phi(x)}$ and 
$g(x)e^{-\phi(x)}$, respectively.

\section{Discussion}

\subsection{Orthogonality for infinitesimal noise}
\label{sec.5.1}

We now consider a general SDE 
with an infinitesimal stochastic term:
\begin{equation}
        dx(t) = b(x)dt + \sqrt{\epsilon} d\xi(t), \ \
        \big\langle\xi(t)\xi(t')\big\rangle
                =2D\delta(t-t'),
\end{equation}
with Fokker-Planck equation
\begin{equation}
  \frac{\partial u(x,t)}{\partial t}
     = -\nabla\cdot J(x,t), \ \
       J(x,t) = b(x)u(x,t)-\epsilon D\nabla u(x,t).
\end{equation}
The large deviation principle suggests that
\cite{fw_70,graham_tel,dykman_ross,ge_qian_ldt}
\begin{equation}
     u^{ss}(x) = \exp\left(-\frac{\psi_{\epsilon}(x)}{\epsilon}\right),
\end{equation}
and thus, $J^{ss}(x) = u^{ss}(x)\big( b(x)+D\nabla\psi_{\epsilon}(x)\big)$.   
Then $\nabla\cdot J^{ss}(x)=0$ implies
\begin{equation}
   \epsilon\nabla\cdot\Big(b(x)+D\nabla\psi_{\epsilon}(x)\Big)=
  \Big( b(x)+D\nabla\psi_{\epsilon}(x)\Big)\cdot\nabla\psi_{\epsilon}(x).
\label{eq_0230}
\end{equation}
Reversibility under time reversal 
$\big(t,\phi,g\big)\rightarrow \big(-t,\phi,g\big)$
is eqivalent to $g=b+D\nabla\psi_{\epsilon}=0$
\cite{jqq,zqq}, which is also known as
potential condition \cite{graham_haken_71,gardiner_book} .
In the present work with time 
reversal $\big(t,\phi,g\big)\rightarrow \big(-t,\phi,-g\big)$,
$g\neq 0$ but
$\nabla\cdot g=g\cdot\nabla\psi_{\epsilon}=0$.
One could consider these conditions as more general
``solvability conditions'' for the steady state of
a stochastic dynamics.

	Eq. \ref{eq_0230} is exact. We now consider the limit 
of small $\epsilon$, and 
$\psi_{\epsilon}(x)=\phi_0(x)+\epsilon\phi_1(x)+\cdots$,
where 
\begin{equation}
       \nabla\phi_0(x)\cdot\big(D\nabla\phi_0(x)+b(x)\big)=0,
\label{eq_0240}
\end{equation}
\begin{equation}
    \nabla\phi_1(x)\cdot\Big(2D\nabla\phi_0(x)+b(x)\Big)
       =\nabla\cdot\Big(D\nabla\phi_0(x)+b(x)\Big),
\label{eq_0250}
\end{equation} 
\begin{equation}
    \nabla\phi_2(x)\cdot\Big(2D\nabla\phi_0(x)+b(x)\Big)
       = \nabla\cdot\Big(D\nabla\phi_1(x)\Big)
           -\big(\nabla\phi_1(x)\big)D\big(\nabla\phi_1(x)\big).
\end{equation}
We thus have a decomposition:
\begin{equation}
  b(x) = -D\nabla\phi_0(x) + \Big(b(x)+D\nabla\phi_0(x)\Big),
\label{eq_0260}
\end{equation}
in which the two terms are orthogonal by the
definition of $\phi_0(x)$, as given in (\ref{eq_0240}). 
This is a well-known result in the theory of large 
deviation \cite{fw_70,ge_qian_ldt}.  The second, however,
is not divergence free.   Actually, Eq. \ref{eq_0230} indicates 
that, in the limit of $\epsilon\rightarrow 0$, 
$\nabla\cdot\big(b(x)+D\nabla\phi_0(x)\big)$ is
$\frac{0}{0}$.  In fact, from (\ref{eq_0250}) 
\begin{equation}
   \nabla\cdot\Big(b(x)+D\nabla\phi_0(x)\Big)
 = -\nabla\phi_1(x)\cdot \Big(2D\nabla\phi_0(x)+b(x)\Big)
\end{equation}
is on the order $O(1)$.  Only when $\phi_1(x)\equiv$
a constant, the equation in (\ref{eq_0260}) 
becomes an orthogonal Helmholtz-Hodge decomposition
in the asymptotic limit of $\epsilon\rightarrow 0$.\footnote{It is intriguing 
to note that according to D. Ruelle, the entropy production of 
a hyperbolic systems (e.g., Anosov diffeomorphisms) 
is zero if its invariant measure has density \cite{ruelle,gallavotti}. 
It only becomes strictly positive when the invariant 
Sinai-Bowen-Ruelle measure is fractal \cite{jqq_cmp}.
One can think of an SBR measure as the limit of a diffusion
process with vanishing $\epsilon$ \cite{ls_young}. 
The present work suggests that equilibrium steady
state with orthogonal $\nabla\phi$ and $g$ guarantees a 
smooth invariant density in the zero-noise limit.  In fact,
one also has a smooth large-deviation rate function 
$\psi_0(x)=\phi(x)$.  Without the orthogonality, the 
invariant measure and the large-deviation rate function 
are in general non-smooth in the zero-noise limit \cite{graham_tel}.
}
Ao's stochastic process has all $\phi_i(x)=0$ 
with $i\ge 1$.

\subsection{Boltzmann's thermodnamic probability 
and Kolmogorov backward equation}

Motivated by Eq. \ref{heat}, let us introduce
$\Omega(x,t)=u(x,t)e^{\phi(x)}$.  Then 
\begin{equation}
  \frac{\partial\Omega(x,t)}{\partial t}
  = \nabla\cdot D\nabla\Omega(x,t) 
    - \Big(g(x)+D\nabla\phi(x)\Big)\cdot\nabla\Omega(x,t).
\label{eq4v}
\end{equation}
The Kolmogorov backward equation corresponding to
(\ref{new_kke}) is 
\begin{equation}
   \frac{\partial v(x,t)}{\partial t}
   = \nabla\cdot D\nabla v(x,t)
                - \Big(D\nabla\phi(x)  
                   -g(x)\Big)
                 \cdot \nabla v(x,t) .
\label{eq4w}
\end{equation}
They differ only by a $g\rightarrow -g$.
The function $\Omega(x,t)$ can be interpreted as 
Boltzmann's ``thermodynamic probability'', whose
logarithm is a thermodynamic potential (Boltzmann 
entropy). Finally,  the probability current consists of a 
``mechanical'' force $g$ and an ``entropic'' force
$-D\nabla\ln\Omega$:
\begin{equation}
   j(q,t) = g(x) - D\nabla\ln\Omega(x,t).
\end{equation}
We also note that for both Eqs. \ref{eq4v} and \ref{eq4w},
if $v(x,t)>0$, there is an Boltzmann's H-theorem like
relation:
\begin{equation}
 \frac{d}{dt}\int_{\mathbb{R}^n} \Big(\Omega(x,t)
        \ln\Omega(x,t)\Big)e^{-\phi(x)}dq
    = -\int_{\mathbb{R}^n} \big(\nabla\Omega\big)\cdot 
            \big(D\nabla\Omega\big) \left(\frac{e^{-\phi}}{\Omega}
                    \right) dx \le 0.
\end{equation}

\subsection{Conservative and dissipative dynamics}

The notions of conserative and dissipative systems
are fundamental concepts in mechanics and 
thermodynamics.  They are the core of  the world 
view based on Newtonian mechanics  \cite{prigogine,ps_bookrev_1,ps_bookrev_2}.
Entropy production is the key mathematical quantity
characterizing dissipation.  Since 1980s, it has
become increasingly clear that the mathematical
foundation of entropy production lies within the 
notion of {\em time reversal}
\cite{jqq,qq_85,crooks,seifert_05,gj_07}.

	For the dynamics associated with
system in (\ref{the_eq}), classical Newtonian notion
of time reversal is 
$\big(t,\phi,g\big)\rightarrow \big(-t,\phi,-g\big)$
which gives rise to an entropy production $e_p^{\#}$ solely 
by the non-adiabatic part $-dF/dt$.
On the other hand, if one chooses time reversal
$\big(t,\phi,g\big)\rightarrow \big(-t,\phi,g\big)$,
then total entropy production $e_p$ is the sum of both 
adiabatic and non-adiabatic entropy 
productions \cite{ge_qian_pre,esposito}.  The 
adiabatic part is also known as house-keeping 
heat \cite{oono_paniconi}: It represents the amount of 
active energy input to sustain a nonequilibrium 
steady state \cite{qian_decomp}.

To put the above discussion into sharper contrast,
consider the following biophysical experiment on 
a single motor protein in the presence of given
ATP and ADP concentrations in solution \cite{chowdhury}.
At 100 cycles per second, a motor protein 
runs for a day with a total $\sim 10^7$ ATP 
hydrolysis.  However, at a millimolar concentration
in a millilitre volume, there are $10^{17}$ total 
number of ATP molecules.  Hence, with a single motor protein
running for a day in such a mesoscopic system, the concentration 
of ATP changes only 1 part of 10 billion.  It is essentially
undetectable.
	
	Now consider an experiment on a type II 
superconducting ring with a current in the 
presence of a magnet. This system has been considered 
in condensed matter physics as an equilibrium system.
However, according to Newton's third law, the 
supercurrent has to exert a force on the magnet
and causing it to slowly demagnetize even though  
it is essentially undetectable \cite{ao_zhu_prb}. 

	Newton's first law states that a linear
constant motion persists in the absence of a force.
A rotatinal motion, hence, requires a force,
even when it does no work.  According to Newton's 
third law and our understanding of the constituents 
of matter, there is always a {\em consequence}, 
{\em \`{a} la} Lord Kelvin, at the origin of 
the force field that causes the rotational motion, 
i.e., a stationary current.  The {\em environment} of a 
system with sustained rotational motion, therefore, can 
not be absolutely time reversible.

	Conservative or dissipative thermodynamic description of an 
open subsystem are mathematical models which depend 
upon an experimentalist's 
knowledge and perspective.  They are rather 
theoretical issues; the dynamics is closer to the 
reality.

\vskip 0.5cm\noindent

I thank Ping Ao, Dick Bedeaux, Zhen-Qing Chen, Mark Dykman, 
Massimiliano Esposito, Hao Ge, Da-Quan Jiang, Signe Kjelstrup, 
David Lacoste, H.-C. \"{O}ttinger, Masaki Sasai, Udo Seifert, 
Jin Wang, and Jianhua Xing for many helpful discussions.
I also thank Professor Madeleine Dong (University of Washington) for 
continuous dialogues on the nature of interpretations, narratives, 
and discourse in historical research.

\end{document}